\documentclass[aps,preprint,epsfig]{revtex4}
\usepackage{graphicx,latexsym,amsmath}
\begin{document}
\title{Free energy and extension of a semiflexible\\ polymer 
in cylindrical confining geometries}
\author{
Yingzi Yang,$^1$ Theodore W. Burkhardt,$^{1,2}$ and Gerhard Gompper$^1$ }
\affiliation{$^1$ Institut f\"ur Festk\"orperforschung, Forschungszentrum
J\"ulich, D-52425 J\"ulich, Germany \\
$^2$ Department of Physics, Temple University, Philadelphia, PA
19122, USA}
\begin{abstract}
We consider a long, semiflexible polymer, with persistence length $P$ and contour length $L$,
fluctuating in a narrow cylindrical channel of diameter $D$. In the regime $D\ll P\ll L$ the free energy of confinement $\Delta F$ and the length of the channel $R_\parallel$ occupied by the polymer are given by Odijk's relations $\Delta F/R_\parallel=A_\circ k_BTP^{-1/3}D^{-2/3}$ and $R_\parallel=L\left[1-\alpha_\circ(D/P)^{2/3}\right]$, where $A_\circ$ and $\alpha_\circ$ are dimensionless amplitudes. Using a simulation algorithm inspired by PERM (Pruned Enriched Rosenbluth Method), which yields results for very long polymers, we determine $A_\circ$ and $\alpha_\circ$ and the analogous amplitudes for a channel with a rectangular cross section. For a semiflexible polymer confined to the surface of a cylinder, the corresponding amplitudes are derived with an exact analytic approach. The results are relevant for interpreting experiments on biopolymers in microchannels or microfluidic devices.
\end{abstract}
\pacs{PACS}
\maketitle

\section{Introduction}
Microfluidic devices provide new possibilities for studying biological polymers such as DNA, actin filaments, and microtubules. Since the persistence lengths of biological polymers are typically tens of nanometers or larger, their behavior in confinement, as in nano- or microchannels, is different from that of flexible synthetic macromolecules.

In this paper we consider the equilibrium statistics of a semiflexible polymer or worm-like chain with persistence length $P$ and contour length $L$ in a channel of diameter $D$. In the regime $D\ll P\ll L$, corresponding to a long, tightly confined polymer, Odijk \cite{to} showed that the free energy of confinement $\Delta F$, i.e. the work required to reversibly insert the polymer in the channel, and the length of the channel $R_\parallel$ occupied by the polymer are given by
\begin{eqnarray} 
&&{\Delta F\over R_\parallel}=A_\circ{k_BT\over P^{1/3}D^{2/3}}\;,\label{DeltaFcirc}\\
&&R_\parallel=L\left[1-\alpha_\circ\left({D\over P}\right)^{2/3}\right]\;.\label{Rparallelcirc}
\end{eqnarray}
For a channel with a rectangular cross section with edges $D_x$ and $D_y$,
\begin{eqnarray} 
&&{\Delta F\over R_\parallel}=A_\Box{k_BT\over P^{1/3}}\left({1\over D_x^{2/3}}+{1\over D_y^{2/3}}\right)\;,\label{DeltaFBox}\\
&&R_\parallel=L\left(1-\alpha_\Box{D_x^{2/3}+D_y^{2/3}\over P^{2/3}}\right)\;.\label{RparallelBox}
\end{eqnarray}
Here $A_\circ$, $\alpha_\circ$, $A_\Box$, and $\alpha_\Box$ are dimensionless universal numbers, which do not depend on $P$, $D$, $D_x$, and $D_y$.

Making use of advances in the manipulation of single polymers, recent experiments have begun to approach the Odijk regime $D\ll P\ll L$. In the experiments of Reisner et al. \cite{retal} on single DNA molecules with persistence length $P$ of about 50 nm and contours lengths $L$ of around 20 $\mu$m or larger, the condition $P\ll L$ is well satisfied, and the dimensions $D_x=30$ nm, $D_y=40$ nm of the narrowest channels are moderately smaller than $P$. In the experiments of K\"oster et al. \cite{ketal} on actin filaments with persistence length of about 20 $\mu$ in microchannels with diameters down to 1 or 2 $\mu$m, $D\ll P$ for the narrowest channels, and the longest contour lengths $L$ considered of around 50 $\mu$m are about 2 to 3 times $P$. For an experiment in which DNA is confined by a grooved substrate instead of a channel, see Hochrein et al. \cite{hlgr}.

For interpreting such experiments it is important to know the numerical values of the dimensionless amplitudes in Eqs. (\ref{DeltaFcirc})-(\ref{RparallelBox}). Solving an integral equation numerically which arises in an exact analytic approach, Burkhardt \cite{twb97} found
\begin{equation}
A_\Box=1.1036\;,\label{ABoxtwb}
\end{equation}
and from simulations Bicout and Burkhardt \cite{bb} obtained
\begin{equation}
A_\Box=1.108\pm 0.013\;,\quad A_\circ=2.375\pm 0.013\;.\label{AcircBoxbb}
\end{equation}
Other estimates from simulations, compatible with these values but with larger error bars, are given in Refs. \cite{dfl,wg,cs}, and related results for a helical polymer in a cylindrical channel in Ref. \cite{lbg}.

Although the free energy amplitudes $A_\Box$ and $A_\circ$ are known with good precision, comparable estimates of the extension amplitudes $\alpha_\Box$, $\alpha_\circ$ have not been available. Thus, we have determined $\alpha_\Box$ and $\alpha_\circ$ from simulations and obtained new, more precise estimates of $A_\Box$ and $A_\circ$, as described in this paper.

The surface of a channel can be prepared so that biopolymers are adsorbed. For example, naturally anionic DNA strands are adsorbed on a surface coated with cationic lipid membranes and have a high lateral mobility on the surface \cite{hlgr,mr1,mr2}. The attractive interaction between the surface and the biopolymer lowers the free energy barrier for insertion of a macromolecule in a narrow channel. With this as motivation we also consider the free energy and extension of a semiflexible polymer confined to the surface of a cylinder.

The theoretical framework for our calculations is outlined in Section II. In Section III a simulation algorithm inspired by PERM (Pruned Enriched Rosenbluth Method) \cite{pg,gfn,gn} is described, which enables us to consider polymers two or more orders of magnitude longer than in the simulations of Ref. \cite{bb}, on which the results (\ref{AcircBoxbb}) are based. Our estimates of the amplitudes $A_\Box$, $A_\circ$, $\alpha_\Box$, and $\alpha_\circ$ are given in Section III. In Section IV we consider a semiflexible  polmer confined to the surface of a cylinder with diameter $D$. In the regime $D\ll P\ll L$ the confinement free energy and extension are also given by Eqs. (\ref{DeltaFcirc}) and (\ref{Rparallelcirc}), but with different amplitudes $A_{\cal S}$, $\alpha_{\cal S}$. These amplitudes are calculated with an exact analytic approach. In the concluding Section V we compare the results of Sections III and IV with predictions for a polymer confined by an effective parabolic potential.

\section{Theoretical framework}
In the worm-like chain model of a semiflexible polymer, the bending energy is given by
\begin{equation}
{\cal H}={\kappa\over 2}\int_0^L ds\left({d\hat{\tau}\over ds}\right)^2\;.\label{wormlikechain}
\end{equation}
Here $\hat{\tau}$ is the unit vector tangent to the polymer contour, $s$ is the arc length, and $\kappa$ is the bending rigidity, related to the persistence length by $P=\kappa/k_BT$.  In the regime $D\ll P\ll L$, backfolding of the polymer and excluded volume effects are negligible. In typical polymer configurations the tangent vector is nearly parallel to the symmetry axis of the channel. The configurations correspond to single valued functions $\vec{r}(t)$, where $(x,y,t)$ are Cartesian coordinates (see Fig. 1), and $\vec{r}=(x,y)$ specifies the transverse displacement of the polymer from the symmetry axis or $t$ axis of the channel. Since $\vert\vec{v}\vert\ll 1$, where $\vec{v}=d\vec{r}/dt$, the bending energy (\ref{wormlikechain}) simplifies to 
\begin{equation}
{\cal H}={\kappa\over 2}\int_0^L dt\left({d^2\vec{r}\over dt^2}\right)^2\;,\label{chaininchannel}
\end{equation}
and the length of the channel $R_\parallel$ occupied by the polymer and the contour length $L$ are related by
\begin{equation}
L=\int_0^{R_\parallel}dt\left(1+\vec{v}^{\;2}\right)^{1/2}\approx R_\parallel+{1\over 2}\int_0^{R_\parallel}dt\;\vec{v}^{\;2}\;.\label{extension1}
\end{equation}

In accordance with Eq. (\ref{chaininchannel}), the partition function of a polymer with position and slope $\vec{r}_0,\vec{v}_0$ at $t=0$ and $\vec{r},\vec{v}$ at $t$ is given by the path integral
\begin{equation}
Z(\vec{r},\vec{v};\vec{r}_0,\vec{v}_0;t)=\int D^2r\;\exp\left[-{P\over 2}\int_0^t dt\left({d^2\vec{r}\over dt^2}\right)^2\right]\;,\label{partitionfunction1}
\end{equation}
where $\vec{r}$ is restricted to the interior of the channel. It satisfies the Fokker-Planck type differential equation
\begin{equation}
\left({\partial\over\partial t}+\vec{v}\cdot{\bf\nabla_r}-{1\over 2P} \nabla_{\bf
v}^2\right)Z(\vec{r},\vec{v};\vec{r}_0,\vec{v}_0;t) =0 \label{fpeqchannel}\;,
\end{equation}
with the initial condition $Z(\vec{r},\vec{v};\vec{r}_0,\vec{v}_0;0)=\delta(\vec{r}-\vec{r}_0)\delta(\vec{v}-\vec{v}_0)$.

The boundary condition at a ``hard'' channel wall follows from the fact that 
that discontinuities in the slope of the polymer cost an infinite bending energy and are suppressed. Thus, as $\vec{r}$ approaches the channel wall, $Z(\vec{r},\vec{v};\vec{r}_0,\vec{v}_0;t)$ vanishes for $\hat{n}\cdot\vec{v}>0$, but not for $\hat{n}\cdot\vec{v}<0$, where $\hat{n}$ is normal to the wall and directed toward the interior of the channel \cite{twb97}.

Our reason for denoting the Cartesian coordinates by $(x,y,t)$ instead of $(x,y,z)$ is explained in Fig. 1. Each polymer configuration $\vec{r}(t)$ may be interpreted as the position of a randomly accelerated particle in two dimensions, plotted as a function of the time $t$. The polymer partition function (\ref{partitionfunction1}) corresponds to the propagator or probability density for propagation from initial position and velocity $\vec{r}_0,\vec{v}_0$ to $\vec{r},\vec{v}$ in a time $t$. From the Boltzmann factor in Eq. (\ref{partitionfunction1}) one sees that the acceleration of the particle at each instant is an independent, Gaussian-distributed random variable, with
\begin{equation}
{d^2\vec{r}\over dt^2}=\vec{\eta}(t)\;,\quad\langle\vec{\eta}(t)\rangle=0\;,\quad
\langle\eta_i(t)\eta_j(t')\rangle={\delta_{ij}\over P}\thinspace\delta(t-t')\;.\label{eqmo}
\end{equation}

Since the polymer partition function vanishes at a hard wall for $\hat{n}\cdot\vec{v}>0$, the propagator for the randomly accelerated particle vanishes if the particle is reflected toward the interior of the two dimensional domain representing the channel cross section. Thus, the hard wall in the polymer problem corresponds to an absorbing boundary for the randomly accelerated particle.

For large $t$ the partion function (\ref{partitionfunction1}) decays as 
\begin{equation}
Z(\vec{r},\vec{v};\vec{r}_0,\vec{v}_0;t)\approx \psi_0(\vec{r},\vec{v})\psi_0(\vec{r}_0,-\vec{v}_0)e^{-E_0(P,D)t} \;,\quad t\to\infty\;,\label{Zlarget}
\end{equation}
where $E_0(P,D)$ is the smallest eigenvalue of the $t$ independent Fokker-Planck equation
\begin{equation}
\left[-E(P,D)+\vec{v}\cdot{\bf\nabla_r}-{1\over 2P} \nabla_{\bf
v}^2\right]\psi(\vec{r},\vec{v})=0 \label{fpeqchanneltindep}\;.
\end{equation}
Together with the definition 
\begin{equation}
\exp\left(-{\Delta F\over k_BT}\right)={Z(D)\over Z(\infty)}\label{confinementfreeenergy}
\end{equation}
 of the free energy of confinement, Eq. (\ref{Zlarget}) implies
\begin{equation}
{\Delta F\over k_BT\;R_\parallel}=E_0^\circ(P,D)-E_0^\circ(P,\infty)={E_0^\circ({1\over 2},1)\over(2P)^{1/3}D^{2/3}}\label{DeltaF2circ}
\end{equation} 
for a channel with a circular cross section, in agreement with Odijk's relation (\ref{DeltaFcirc}). Here we have used the scaling relation $E_0^\circ(P,D)=(2P)^{-1/3}D^{-2/3}E_0^\circ({1\over 2},1)$, which is readily derived by rewriting Eq. (\ref{fpeqchanneltindep}) in terms of the dimensionless variables $\vec{r}\;'=D^{-1}\vec{r}$, $t'=(2P)^{-1/3}D^{-2/3}t$, $\vec{v}\;'=(2P)^{1/3}D^{-1/3}\vec{v}$.

For a channel with a rectangular cross section with edges $D_x$, $D_y$, the partition function in Eq. (\ref{partitionfunction1}) has the product form $Z(\vec{r},\vec{v};\vec{r}_0,\vec{v}_0;t)=Z(x,v_x;x_0,v_{x0};t)Z(y,v_y;y_0,v_{y0};t)$. This is the origin of the sum of independent $x$ and $y$ contributions in Eqs. (\ref{DeltaFBox}) and (\ref{RparallelBox}). The solutions to Eq. (\ref{fpeqchanneltindep}) also have the separable form $\psi(\vec{r},\vec{v})=\psi(x,v_x,)\psi(y,v_y)$, implying $E_0^\Box(P,D_x,D_y)=E_0^\vert(P,D_x)+E_0^\vert(P,D_y)$. Here $E_0^\vert(P,D_x)$ is the smallest eigenvalue of the equation
\begin{equation}
\left[-E^\vert(P,D)+v_x{\partial\over\partial x}-{1\over 2P}{\partial^2\over\partial v_x^2}
\right]\psi(x,v_x)=0 \label{fpeqchanneltindeponedim}\;.
\end{equation}
on the one-dimensional interval $-{1\over 2}D_x <x< {1\over 2}D_x\;$, with boundary condition $\psi(-{1\over 2}D_x,v_x)=\psi({1\over 2}D_x,-v_x)=0$ for $v_x>0$. The scaling relation $E_0^\vert(P,D_x)=(2P)^{-1/3}D_x^{-2/3}E_0^\vert({1\over 2},1)$, is readily derived by rewriting Eq. (\ref{fpeqchanneltindeponedim}) in terms of the dimensionless variables $x'=D^{-1}x$, $t'=(2P)^{-1/3}D_x^{-2/3}t$, $v_x'=(2P)^{1/3}D_x^{-1/3}v_x$.

From Eqs. (\ref{DeltaFcirc}), (\ref{DeltaFBox}), (\ref{DeltaF2circ}), and the results of the preceding paragraph, we obtain
\begin{equation}
A_\circ=2^{-1/3}E_0^\circ({\textstyle{1\over 2}},1)\;,\quad A_\Box=2^{-1/3}E_0^\vert({\textstyle{1\over 2}},1)\;.\label{AcircBox}
\end{equation}

To obtain comparable expressions for the amplitudes $\alpha_\circ$, $\alpha_\Box$, we begin by comparing Eqs. (\ref{Rparallelcirc}) and (\ref{extension1}), which imply 
\begin{equation}
\alpha_\circ={1\over 2}\left({P\over D}\right)^{2/3}\langle\vec{v}^{\;2}\rangle_{P,D}^\circ\;.\label{alphacirc1}
\end{equation}
Here $\langle\vec{v}^{\;2}\rangle_{P,D}^\circ$ is the average value of $\vec{v}^{\;2}$ along an infinitely long, tightly confined polymer in a channel with a circular cross section. In terms of the ground state eigenfunction $\psi_0(\vec{r},\vec{v})$ of Eq. (\ref{fpeqchanneltindep}) with eigenvalue 
$E_0^\circ(P,D)$,
\begin{equation}
\langle\vec{v}^{\;2}\rangle_{P,D}^\circ\;=\;{\int d^2r\int d^2v \;\vec{v}^{\;2}\psi_0(\vec{r},\vec{v})\psi_0(\vec{r},\vec{-v})\over \int d^2r\int d^2v \;\psi_0(\vec{r},\vec{v})\psi_0(\vec{r},\vec{-v})}\;.\label{alphacirc2}
\end{equation}
Expressing Eq. (\ref{alphacirc1}) and its analog for the rectangular cross section in terms of the dimensionless variables introduced below Eqs. (\ref{DeltaF2circ}) and (\ref{fpeqchanneltindeponedim}), we obtain
\begin{equation}
\alpha_\circ=2^{-5/3}\langle\vec{v}\;'^2\rangle_{{1\over 2},1}^\circ\;\;,\quad
\alpha_\Box=2^{-5/3}\langle v_x'^2\rangle_{{1\over 2},1}^{\;\vert}\;\;,\label{alphacircBox}
\end{equation}

Equations (\ref{AcircBox}) and (\ref{alphacircBox}) play a central role in our work, allowing us to determine the free energy and extension amplitudes from simulations with $P={1\over 2}$ and $D=D_x=D_y=1$.

\section{Simulations} 
\subsection{Algorithm}
To determine $A_\Box$ and $\alpha_\Box$ from simulations, we generate a large number $N_0$ of configurations $x'(t')$ of a polymer with persistence length $P={1\over 2}$ in the {\em unbounded} two-dimensional space $(x',t')$. Here $x'$ and $t'$ are the dimensionless coordinates introduced below Eq. (\ref{fpeqchanneltindeponedim}). The configurations are generated with same Boltzmann weight as in Eq. (\ref{partitionfunction1}), but in two rather than three spatial dimensions. All of the configurations have the same initial position and slope $x'_0=v'_0=0$ at $t'_0=0$. Each configuration is ``grown'' until it leaves the interval $-{1\over 2}<x'<{1\over 2}$ for the first time. From this information we calculate the fraction $Q(t')$ of the $N_0$ configurations which have not yet left the interval at $t'$.

From Eq. (\ref{confinementfreeenergy}) we see that $Q(t')=\exp(-\Delta F/k_BT)$, where $\Delta F$ is the free energy of confinement of a polymer with one end fixed, as described above, which extends a distance $t'$ down a two-dimensional channel of width 1. According to Eq. (\ref{Zlarget}) and the discussion below Eq. (\ref{fpeqchanneltindeponedim}), $Q(t')$ decays as 
\begin{equation}
Q(t')\sim e^{-E_0^\vert({1\over 2},1)t'}\label{expodecay}
\end{equation}
for large $t'$. To estimate $A_\Box$, we fit the $Q(t')$ extracted from the simulations with the exponential form (\ref{expodecay}) for large $t'$ to obtain $E_0^\vert({1\over 2},1)$ and then use Eq. (\ref{AcircBox}).

The polymer configurations are generated with the algorithm 
 \begin{eqnarray}
&&x'_{n+1}=x'_n+v'_n\Delta_{n+1}+\left({\Delta_{n+1}^3\over 6}\right)^{1/2}\left(s_{n+1}
+\sqrt 3\thinspace r_{n+1}\right)\;,\label{xalgo}\\
&&v'_{n+1}=v'_n+\left(2\Delta_{n+1}\right)^{1/2}r_{n+1}\;,\label{valgo}
\end{eqnarray}
introduced in Ref. \cite{bb1st} and also used in Refs. \cite{bb,kb05}.
Here $x'_n$ is the position of the polymer at point $t'_n$, and $\Delta_{n+1}=t'_{n+1}-t'_n$ is the length step. The quantities $r_n$ and $s_n$ are independent, Gaussian random numbers with $\langle r_n\rangle=\langle s_n\rangle=0$ and $\langle r_n^2\rangle=\langle s_n^2\rangle=1$.

As discussed in Refs. \cite{bb,bb1st,kb05}, this algorithm generates polymer configurations consistent with the Boltzmann weight (\ref{partitionfunction1}) in free space, i.e., in the absence of boundaries. An advantage of the algorithm is that in free space there is no length-step error. The length step $\Delta_{n+1}$ need not be small. For good efficiency we use a fairly large step when $x'_n$ is well inside the interval $-{1\over 2}<x'<{1\over 2}$. Near $x'=\pm{1\over 2}$ a smaller step is needed in order to accurately determine the value of $t'$ at which the configuration leaves the interval for the first time, and hence $Q(t')$. As in Ref. \cite{bb}, we choose 
\begin{equation}
\Delta_{n+1}=10^{-1}\left({\textstyle{1\over 2}}-\vert x\vert\right)+10^{-5}\;,\label{Deltat}
\end{equation}
which varies from 0.05 at $x'=0$ to $10^{-5}$ at $x'=\pm{1\over 2}$.
Further reduction of the length step had no significant effect on our estimates.

In estimating $\langle v_x'^2\rangle_{{1\over 2},1}^{\;\vert}$ in Eq. (\ref{alphacircBox}) to determine $\alpha_\Box$, one should only use the subset of the configurations, generated as described above, which lie entirely within the interval $-{1\over 2}<x'<{1\over 2}$, i.e., within the channel. For each of the configurations ${\cal C}$ in the subset we average $v'^2$, over the entire length of each of the configurations in the subset, using 
\begin{equation}
\langle v'^2\rangle_{\cal C}=
\sum_n{\Delta_{n+1}\over t'}\langle v'^2\rangle_{n+1}\;.\label{vsq1}
\end{equation}
Here $t'=\sum_n\Delta_{n+1}$ is the total length of the channel occupied by configuration ${\cal C}$, and $\langle v'^2\rangle_{n+1}$
is the equilibrium value of $v'^2$ for a semiflexible polymer with endpoints $(x'_n,v'_n)$ and $(x'_{n+1},v'_{n+1})$ at $t'_n=$ and $t'_{n+1}$, respectively, averaged over all intermediate $t'$. This quantity is readily calculated from the free space partition function or propagator and is given by
\begin{eqnarray}
\langle v'^2\rangle_{n+1}&=&{\textstyle{2\over 15}}\Delta_{n+1}+{\textstyle{6\over 5}}\left(x'_{n+1}-x'_n\right)^2\Delta_{n+1}^{-2}
+{\textstyle{2\over 15}}\left(v_{n+1}'^2-{\textstyle{1\over 2}}v_{n+1}'v_n'+v_n'^2\right)\nonumber\\&&\quad
-{\textstyle{1\over 5}}\left(x'_{n+1}-x'_n\right)\left(v'_{n+1}+v'_n\right)\Delta_{n+1}^{-1}\;.\label{vsq2}
\end{eqnarray}
Having calculated $\langle v'^2\rangle_{\cal C}$ for each configuration in the subset this way, we average the results over all the  configurations in the subset to obtain an estimate of $\langle v_x'^2\rangle_{{1\over 2},1}^{\;\vert}$ and, using Eq. (\ref{alphacircBox}), the corresponding value of $\alpha_\Box$.

\subsection{Enrichment Procedure}
The quantity $\langle v_x'^2\rangle_{{1\over 2},1}^{\;\vert}$ in Eq. (\ref{alphacircBox}) is the average value of $v^2$ for a semiflexible polymer of {\em infinite} length in a channel. We found it necessary to go to lengths $t'$ of around 100 to estimate $\langle v_x'^2\rangle_{{1\over 2},1}^{\;\vert}$, free of finite-length effects, to 3 significant figures. However, it is not feasible to generate configurations this long, which lie entirely within the channel, without modifying the steps outlined in the preceding paragraphs. From Eqs. (\ref{ABoxtwb}) and (\ref{alphacircBox}), $E_0({1\over 2},1)$ is close to $1.390$. Thus, according to Eq. (\ref{expodecay}), the probability that a configuration of length t'=100, generated as described above, never leaves the channel $-{1\over 2}<x'<{1\over 2}$, is about $e^{-139}\approx 10^{-61}$.

To generate a large, statistically useful number of configurations lying entirely in the channel, we used an enrichment procedure inspired by PERM (Pruned Enriched Rosenbluth Method) \cite{pg,gfn,gn}, which has been successfully applied in simulations of a wide variety of systems, including flexible, self-avoiding polymers in channels \cite{fcg,hg}. 

We begin by generating a large number $N_0$ of configurations as described above. Let $N_1$ be the number of these configurations which have not yet left the channel at $t'=\tau$. We make $n$ copies of each of these configurations and then, with the algorithm of the preceding Subsection, continue each of the $n N_1$ configurations past $t'=\tau$. Let $N_2$ be the number of these configurations which have not yet left the channel at $t'=2\tau$. Again we make $n$ copies and then continue the $nN_2$ configurations past $t'=2\tau$. At $t'=3\tau,4\tau ,\dots$ the same procedure is followed.

To estimate $A_\Box$ using Eqs. (\ref{AcircBox}) and (\ref{expodecay}), we need to calculate the probability $Q(t')$, defined above Eq. (\ref{expodecay}), that a configuration, generated as in the preceding Subsection, has not yet left the channel at $t'$. To obtain this probability, it is useful to think of copying all the preceding configurations, including the number of initial configurations, at $t'=\tau,2\tau. \dots$. Thus,
\begin{equation}
Q(0)=1\;,\quad Q(k\tau)=
{N_k\over n^{k-1}N_0}\;\;{\rm for}\;k=1,2,\dots\label{Q(k)}
\end{equation}

Our results for a semiflexible polymer in two dimensions were obtained with $N_0=1.8\times 10^7$,
$\tau=1$, and $n=4$. These values of $n$ and $\tau$ were chosen so that $N_k$ slowly decreases with increasing $k$. From Eqs. (\ref{ABoxtwb}), (\ref{AcircBox}), (\ref{expodecay}), and (\ref{Q(k)}), one finds $N_k\sim n^ke^{-E_0^\vert({1\over 2},1)k\tau}N_0=(0.996)^k N_0$\;.

We have also calculated the number of families $N_k^{\rm fam}$ to which the $N_k$ configurations that remain in the channel up to $t'=k\tau$ belong. Two configurations are said to belong to the same family if they coincide in the interval $0<t'<\tau$, i.e., if their most remote ancestor is the same. By definition  $N_1^{\rm fam}=N_1$, but for larger $k$,
$N_k^{\rm fam}\leq N_k$, since several of the $N_k$ configurations may belong to the same family. According to our simulation data  $N_k^{\rm fam}$ also decays as $n^ke^{-E_0({1\over 2},1)k\tau}=(0.996)^k$. For sufficiently large $k$ all $N_k$ configurations belong to a single family.

To determine $\alpha_\Box$, we evaluate $\langle v_x'^2\rangle_{{1\over 2},1}^{\;\vert}$, as outlined above in the paragraph containing Eqs. (\ref{vsq1}) and (\ref{vsq2}), for those $N_k$ configurations which remain in the channel up to $t'=k\tau$, estimate the limiting value for large $t'$, and then use Eq. (\ref{alphacircBox}).

The simulations of a polymer in a channel with a circular cross section of diameter $D$ are very similar. In terms of the dimensionless Cartesian coordinates $(x',y',t')$ introduced below Eq. (\ref{DeltaF2circ}), the channel has radius ${1\over 2}$. Beginning with $x'_0=y'_0=v_{x0}'=v_{y0}'=t'_0=0$, we generate the sequence $(x'_n,y'_n,t'_n)$ with the algorithm (\ref{xalgo})-(\ref{valgo}) and corresponding equations with $x$ replaced by $y$. In analogy with Eq. (\ref{Deltat}) the length step is 
\begin{equation}
\Delta_{n+1}=10^{-1}\left[{\textstyle{1\over 2}}-\left(x_n'^2+y_n'^2\right)^{1/2}\right]+10^{-5}\;.\label{Deltatcirc}
\end{equation}

Each configuration is grown until it leaves the circular domain $\left(x'^2+y'^2\right)^{1/2}<{1\over 2}$. Again we begin with $N_0$ configurations and at $t'=\tau,2\tau,\dots$ make $n$ copies of the $N_1,N_2,\dots$, configurations which have not yet left the circular domain. Our results were obtained with $N_0=4.7\times 10^7$,
$\tau=1.009$, and $n=20$. As in the two-dimensional case these parameters were chosen so that $N_k$ and $N_k^{\rm fam}$ decay rather slowly, as $n^ke^{-E_0^\circ({1\over 2},1)k\tau}=(0.997)^k$, where we have used Eqs. (\ref{AcircBox}), (\ref{expodecay}), (\ref{Q(k)}), and our result for $A_\circ$ in Eq. (\ref{ABoxcirc}).

To estimate  $A_\circ$, we calculate $Q(t')$ for integer $t'$ using Eq. (\ref{Q(k)}), fit the results with the exponential form (\ref{expodecay}), but with $E_0^\circ({1\over 2},1)$ in place of $E_0^\vert({1\over 2},1)$, and then use Eq. (\ref{AcircBox}). To estimate $\alpha_\circ$, we evaluate $\langle\vec{v}\;'^2\rangle_{{1\over 2},1}^\circ$, as described in connection with Eqs. (\ref{vsq1}) and (\ref{vsq2}), for those $N_k$ configurations which remain in the channel up to $t'=k\tau$ and then use Eq. (\ref{alphacircBox}). 

\subsection{Results}
In Fig. 2, $\ln Q(t')$, as determined from Eq. (\ref{Q(k)}), is shown for $t'=0,\;\tau,\;2\tau,\dots$ for a polymer on a two-dimensional strip (upper curve) and in a three-dimensional channel with a circular cross section (lower curve). According to Eqs. (\ref{expodecay}) and (\ref{AcircBox}) the upper and lower curves have slope $E_0^\vert({1\over 2},1)=2^{1/3}A_\Box$ and $E_0^\circ({1\over 2},1)=2^{1/3}A_\circ$, respectively, for large $t'$. From the best fit to the slope, we obtain
\begin{equation}
A_\Box=1.1038\pm 0.0006\;,\quad A_\circ=2.3565\pm 0.0004\;.\label{ABoxcirc}
\end{equation}
The uncertainty was estimated by determining $A_\Box$ and $A_\circ$ in each of the intervals  $100<t'<200$, $200<t'<300$,\;...\;, $900<t'<1000$ and quoting a value somewhat larger than the width of the corresponding distribution.

The estimate for $A_\Box$ is in extremely good agreement with the result in Eq. (\ref{ABoxtwb}), obtained by solving an integral equation numerically that determines $A_\Box$ in an exact analytic approach \cite{twb97}. The new estimates for $A_\Box$ and $A_\circ$ in Eq. (\ref{ABoxcirc}) have smaller error bars than the earlier estimates \cite{bb} shown in Eq. (\ref{AcircBoxbb}), which are based on simulations of much shorter polymer chains.

In Fig. 3 the $t'$ dependence of $\langle v_x'^2\rangle_{{1\over 2},1}^{\;\vert}$ and $\langle \vec{v}\;'^2\rangle_{{1\over 2},1}^{\;\circ}$is shown. The averages are based on the configurations which remain in the channel from the starting point up to $t'$. For $t'$ greater than roughly 100 the curves are consistent, within statistical fluctuations, with the constant values $\langle v_x'^2\rangle_{{1\over 2},1}^{\;\vert}=0.2901\pm 0.0003$ and $\langle\vec{v}\;'^2\rangle_{{1\over 2},1}^\circ=0.5400\pm 0.0004$. Substituting these values in Eq. (\ref{alphacircBox}), we obtain
\begin{equation}
\alpha_\Box=0.09137\pm 0.00007\;,\quad\alpha_\circ=0.1701\pm 0.0001\;.\label{alphaBoxcirc}
\end{equation}

For a polymer on a two-dimensional strip, the probability distribution $P(\langle v'^2\rangle)$ of the quantity $\langle v'^2\rangle$ is shown for representative values of $t'$ in Fig. 4. The distribution was determined from the results for $\langle v'^2\rangle_{\cal C}$, where the index ${\cal C}$ labels the configurations that remain in the channel up to length $t'$, and  $\langle v'^2\rangle_{\cal C}$ is the average value of $v'^2$ along configuration ${\cal C}$ from the starting point up to $t'$, calculated as in Eqs. (\ref{vsq1}) and (\ref{vsq2}). The distributions in Fig. 4, are approximately Gaussian, and the half width or standard deviation $w$, shown in Fig. 5, decreases in good agreement with the $t'^{-1/2}$ law expected for statistically independent contributions. Results similar to those in Figs. 4 and 5 were also obtained for a polymer in a channel with a circular cross section.

Since $\langle v'^2\rangle_{\cal C}$ determines the contour length $L'$ of configuration ${\cal C}$ via Eq. (\ref{extension1}), the curves in Fig. 4 may be interpreted as distributions of the contour length $L'$ for fixed $t'$. Presumably the distribution of $t'$ for fixed $L'$, i.e., the end-to-end distribution for a polymer of fixed contour length \cite{lm}, is very similar.
 
\section{Semiflexible polymer confined to a cylindrical surface}
In this Section we consider a semiflexible polymer constrained to lie on a cylindrical surface with a circular cross section. As mentioned in the introduction, this is an obvious model for a semiflexible polymer adsorbed on a channel wall. We analyze the case in which only configurations that leave the polymer in contact with the cylindrical surface are allowed and the equilibrium statistics is determined by the Boltzmann weight $e^{-{\cal H}/k_BT}$, where ${\cal H}$ is the bending energy ({\ref{wormlikechain}). The bending energy is clearly minimized if the polymer configuration is a straight line parallel to the symmetry axis of the channel. In the limit $D\ll P\ll L$ the free energy of confinement and the extension again are given by Eqs.  (\ref{DeltaFcirc}) and (\ref{Rparallelcirc}), as shown below, but with different amplitudes $A_{\cal S}$, $\alpha_{\cal S}$, which we determine.

In the regime $D\ll P\ll L$, typical configurations of a semiflexible polymer correspond to single valued functions $\vec{r}(t)=\hat{r}(t)R$, where $\vec{r}$ specifies the transverse displacement of the polymer from the symmetry axis or $t$ axis of the channel. Here $(r,\theta,t)$ are cylindrical coordinates, and $\hat{r},\hat{\theta},\hat{t}$ are the corresponding unit vectors. Keeping in mind that $d\hat{r}/d\theta=\hat{\theta}$,  $d\hat{\theta}/d\theta=-\hat{r}$, and that $\vert Rd\theta/dt\vert\ll 1$, one finds that the bending energy (\ref{wormlikechain}) takes the form
\begin{equation}
{\cal H}={\kappa\over 2}\int_0^L dt\left[\left(R\;{d^2\theta\over dt^2}\right)^2+
{1\over R^2}\left(R\;{d\theta\over dt}\right)^4\right]\;.\label{chainoncylinder}
\end{equation}

As discussed below Eq. (\ref{fpeqchannel}), the polymer configuration $\vec{r}(t)$ may be interpreted as the position of a randomly accelerated particle in the $(x,y)$ plane, plotted as a function of time. A polymer confined to the surface of a cylinder corresponds to a particle moving on a circle of radius $R$. The first and second terms in the integrand in Eq. (\ref{chainoncylinder}) are the squares of the tangential and centripetal accelerations, respectively.

Equation (\ref{chainoncylinder}) and the definitions $x=R\theta={1\over 2}D\theta$ and $v={1\over 2}D\;d\theta/dt$ lead to the partition function
\begin{equation}
Z(x-x_0,v,v_0,t)=\int Dx\exp\left\{-{P\over 2}\int_0^t dt\left[\left({d^2x\over dt^2}\right)^2+{4\over D^2}\left({dx\over dt}\right)^4\right]\right\}
\label{partitionfunction3}
\end{equation}
and the Fokker-Plack equation
\begin{equation}
\left({\partial\over\partial t}+v{\partial\over\partial x}+{2P\over D^2}\;v^4-{1\over 2P}{\partial^2\over\partial v^2}\right)Z(x-x_0,v,v_0,t) =0 \label{fpeqoncylinder}\;.
\end{equation}
Disregarding the position of the polymer endpoint, we integrate Eq. (\ref{fpeqoncylinder}) over $x$ from $-\infty$ to $\infty$. This yields the Schr\"odinger equation
\begin{equation}
\left({\partial\over\partial t}+{2P\over D^2}\;v^4-{1\over 2P}{\partial^2\over\partial v^2}\right)Z(v,v_0,t) =0 \label{Schreq}\;.
\end{equation}
Equation (\ref{Schreq}) also follows directly from the path integral $Z(v,v_0,t)=\int Dv\exp\left\{-{1\over 2}P\int_0^t dt\left[(dv/dt)^2+(4/D^2) v^4\right]\right\}$, which has the same  Boltzmann factor as in Eq. (\ref{partitionfunction3}), but expressed in terms of $v$ rather than $x$.

Beginning with Eq. (\ref{Schreq}) and following the steps that led from Eq. (\ref{fpeqchannel}) to Eqs. (\ref{AcircBox}) and (\ref{alphacircBox}), we obtain
\begin{equation}
A_{\cal S}=2^{-1/3}E_0^{\cal S}({\textstyle{1\over 2}},1)\;,\quad \alpha_{\cal S}=2^{-5/3} \\
\langle v'^2\rangle_{{1\over 2},1}^{\cal S}\;\;,\label{Aalphaoncylinder}
\end{equation}
Here $E_0^{\cal S}({\textstyle{1\over 2}},1)$ is the ground state energy of the Schr\"odinger equation with a quartic potential
\begin{equation}
\left(-E^{\cal S}+v'^4-{\partial^2\over\partial v'^2}\right)\psi(v') =0 \label{tindepSchreq}\;,
\end{equation}
written in terms of the dimensionless variables introduced below Eq. (\ref{fpeqchanneltindeponedim}), $\psi_0(v')=\psi_0(-v')$
is the wave function of the ground state, and $\langle v'^2\rangle_{{1\over 2},1}^{\cal S}$ is the quantum mechanical expectation value of $v'^2$ in the ground state. Equation ({\ref{Aalphaoncylinder}) allows us to calculate the free energy and extension amplitudes from simulations with $P={1\over 2}$ and $D=1$.

The ground state energy of Eq. (\ref{tindepSchreq}), determined numerically by Bender et al. \cite{bender} and Voros \cite{voros}, is given by $E_0^{\cal S}({\textstyle{1\over 2}},1)=1.060\;362\;09$.
Solving the Schr\"odinger equation (\ref{tindepSchreq}) numerically for this energy using {\it Mathematica}, we obtain $\langle v'^2\rangle_{{1\over 2},1}^{\cal S}=0.362\;023$. Substitution of these values in Eq. (\ref{Aalphaoncylinder}) yields the amplitudes
\begin{equation}
A_{\cal S}=0.84161\;,\quad \alpha_{\cal S}=0.11403\;.\label{Aalphaoncylinder2}
\end{equation}

\section{Concluding remarks}
Using a PERM inspired simulation algorithm, we have determined the amplitudes $A_\Box$, $A_\circ$, $\alpha_\Box$, and $\alpha_\circ$ for a semiflexible polymer in a channel with an estimated error of less than a tenth of a percent. We hope the results will be useful in analyzing experiments. For a polymer confined to the surface of a cylinder, we have calculated the corresponding amplitudes $A_{\cal S}$, $\alpha_{\cal S}$ exactly to five significant figures with an analytical approach. These latter results may be used as a benchmark for testing simulation algorithms.

A common approximation in studies of semiflexible polymers in channels is to replace the hard wall interaction by an effective parabolic potential \cite{lm,twb95}. In concluding, we use this approximation to relate the free energy and extension amplitudes $A$ and $\alpha$ and compare the relations with our simulation results.

The partition function of a polymer which is tightly confined along the $t$ axis by a parabolic potential energy per unit length $u={1\over 2}\thinspace k_BT\thinspace b\vec{r}\;^2$ is given by the path integral
\begin{equation}
Z(\vec{r},\vec{v};\vec{r}_0,\vec{v}_0;t)=\int D^2r\;\exp\left\{-{1\over 2}\int_0^t dt\left[P\left({d^2\vec{r}\over dt^2}\right)^2+b\vec{r}^{\;2}\right]\right\}\;,\label{partitionfunctionparabolic}
\end{equation}
It can be calculated exactly \cite{twb95,lm} and yields
\begin{eqnarray}
&& f(b,P)=k_BT\left({4b\over P}\right)^{1/4}\;,\label{fparabolic}\\
&&\langle\vec{v}^{\;2}\rangle=\left(4b P^3\right)^{-1/4}\;,\label{vsqparabolic}
\end{eqnarray}
where $f=-k_BT\lim_{t\to\infty}t^{-1}\ln Z$ is the free energy per unit length, and $\langle\vec{v}^{\;2}\rangle$ is evaluated in the same limit $t\to\infty$.

For a parabolic potential the confinement free energy per unit length  is $\Delta f=f(b,P)-f(0,P)=f(b,P)$. To obtain a prediction for $\alpha_\circ$, we choose $b$ so that this $\Delta f$ exactly reproduces expression (\ref{DeltaFcirc}) for the free energy of confinement in a channel with a circular cross section, evaluate $\langle\vec{v}^{\;2}\rangle$ in Eq. (\ref{vsqparabolic}) for this $b$, and then substitute the result in Eq. (\ref{alphacirc1}). This and a similar calculation for a channel with a square cross section yield
\begin{equation}
\alpha_\Box A_\Box={1\over 8}\;,\quad\alpha_\circ A_\circ={1\over 2}\;.
\label{parabolicprediction1}
\end{equation}

Ubbink and Odijk \cite{uo} argue that the parabolic confining potential is an artifice, and that the confinement free energy (\ref{DeltaFcirc}) of the polymer in a channel should not be identified with the full free energy $f$ in Eq. (\ref{fparabolic}) but with the configurational part
$f_{\rm config}=f-\langle u\rangle=f-b\partial f/\partial b ={3\over 4}f$.
This leads to \cite{joetal}
\begin{equation}
\alpha_\Box A_\Box={3\over 32}\;,\quad\alpha_\circ A_\circ={3\over 8}\;,
\label{parabolicprediction2}
\end{equation}
which differs from Eq. (\ref{parabolicprediction1}) by an extra factor of ${3\over 4}$ on the right hand side.
Our simulation results in Eqs. (\ref{ABoxcirc}), (\ref{alphaBoxcirc}), and (\ref{Aalphaoncylinder2}) yield the products
\begin{equation}
\alpha_\Box A_\Box=0.1009\pm 0.0002\;,\quad 
\alpha_\circ A_\circ=0.4008\pm 0.0003\;,\quad 
\alpha_{\cal S}A_{\cal S}=0.095969\;, \label{ourproducts}
\end{equation}
which lie in between the predictions (\ref{parabolicprediction1}) and  (\ref{parabolicprediction2}) but in every case closer to (\ref{parabolicprediction2}). Since the surface of a cylinder is two-dimensional, the product $\alpha_{\cal S}A_{\cal S}$ should be compared with the predictions for $\alpha_\Box A_\Box$ in Eqs. (\ref{parabolicprediction1}) and (\ref{parabolicprediction2}), which apply in both two and three dimensions.

\acknowledgments
We thank  Hsiao-Ping Hsu, Walter Nadler, and Roland Winkler for helpful discussions. Hsiao-Ping Hsu showed us how to set up a PERM program for the polymer on the cylinder and has confirmed some of our results. TWB greatly appreciates the hospitality of GG and coworkers at the Forschungszentrum J\"ulich.

\newpage
\begin{figure}[Figure1]
\begin{center}
\includegraphics[width=1.\textwidth,bb=90 460 530 640,clip]{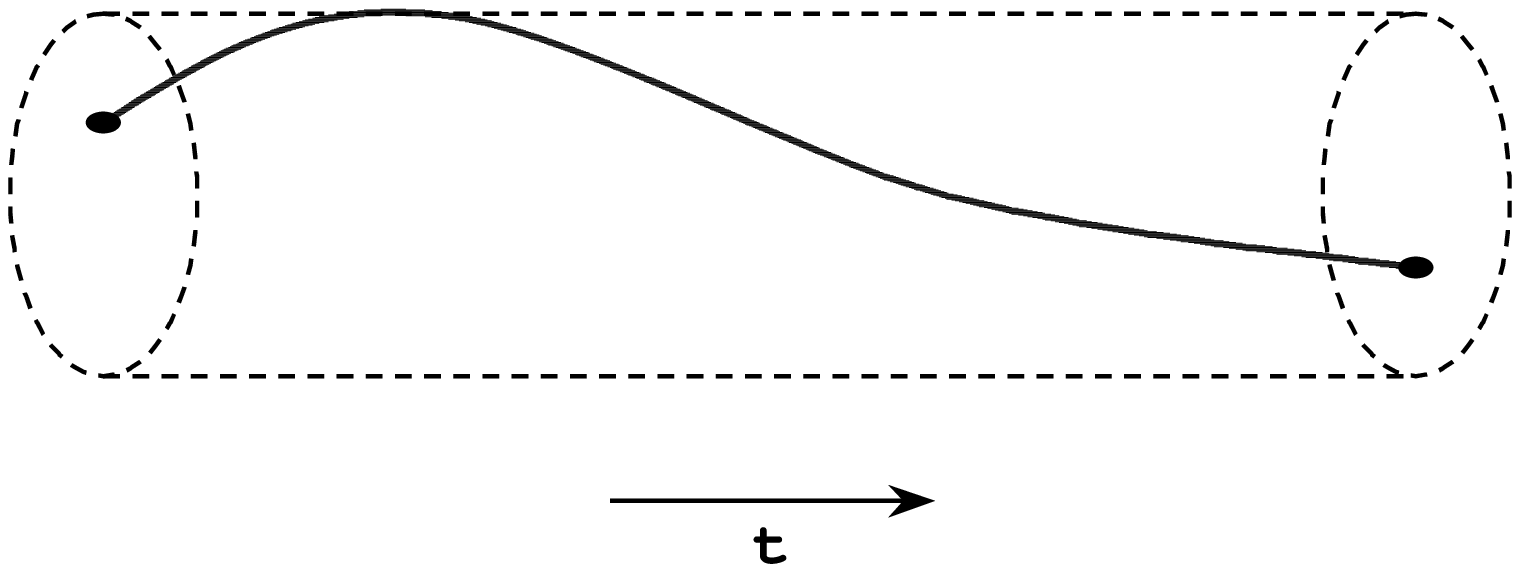}
\caption{The curve may be interpreted as a tightly confined semiflexible polymer in
a channel with a circular cross section or as the trajectory $\vec{r}(t)$, plotted as a
function of $t$, of a randomly accelerated particle moving in two dimensions, which
has not yet left a circular domain.} \label{fig1}
\end{center}
\end{figure}

\newpage
\begin{figure}[Figure2]
\begin{center}
\includegraphics[width=18cm]{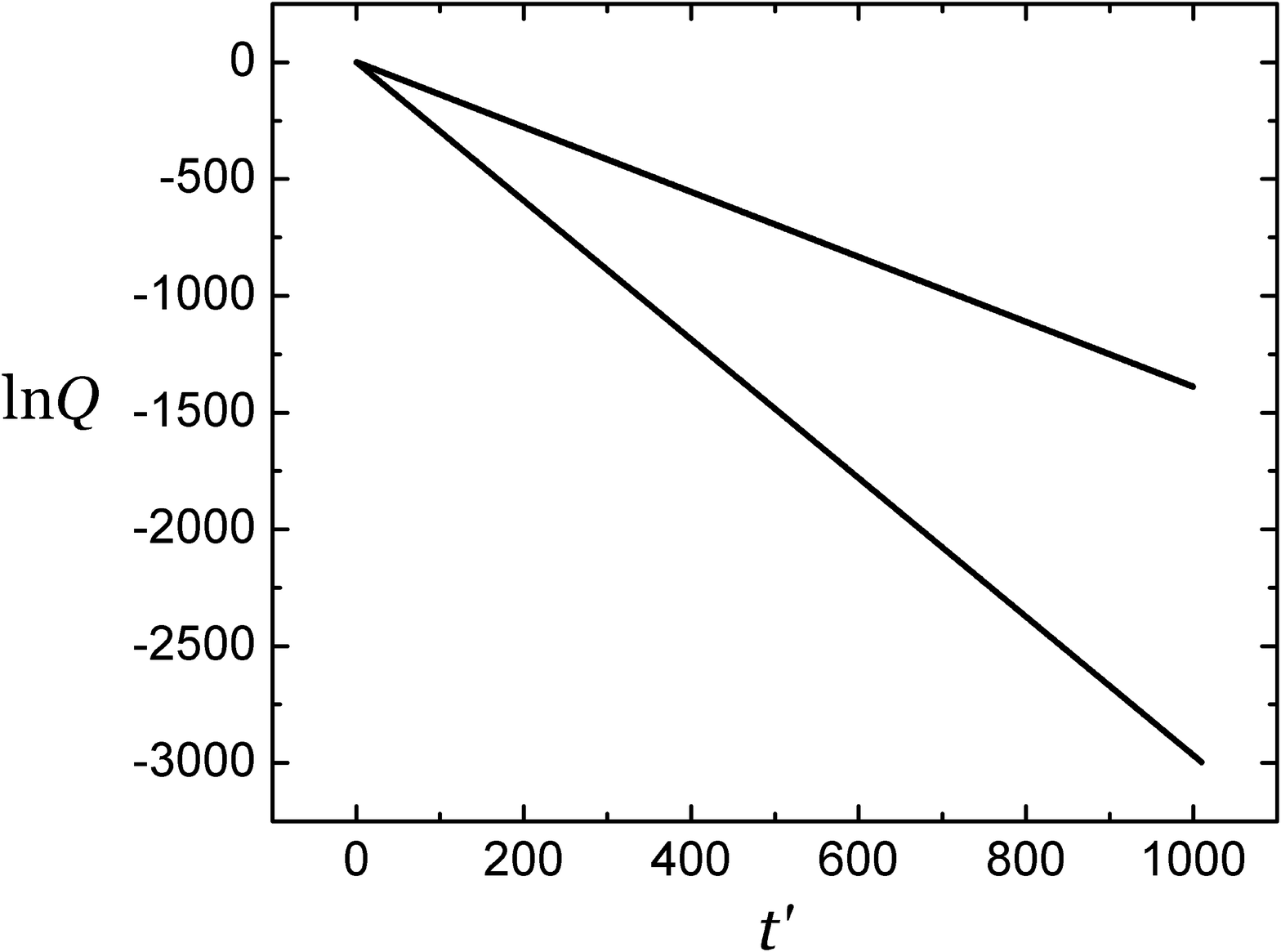}
\caption{$\ln Q(t')$ vs. $t'$ for a polymer on a two-dimensional strip (upper curve) and for a polymer in a three-dimensional channel with a circular cross section (lower curve).} \label{fig2}
\end{center}
\end{figure}

\newpage
\begin{figure}[Figure3]
\begin{center}
\includegraphics[width=18cm]{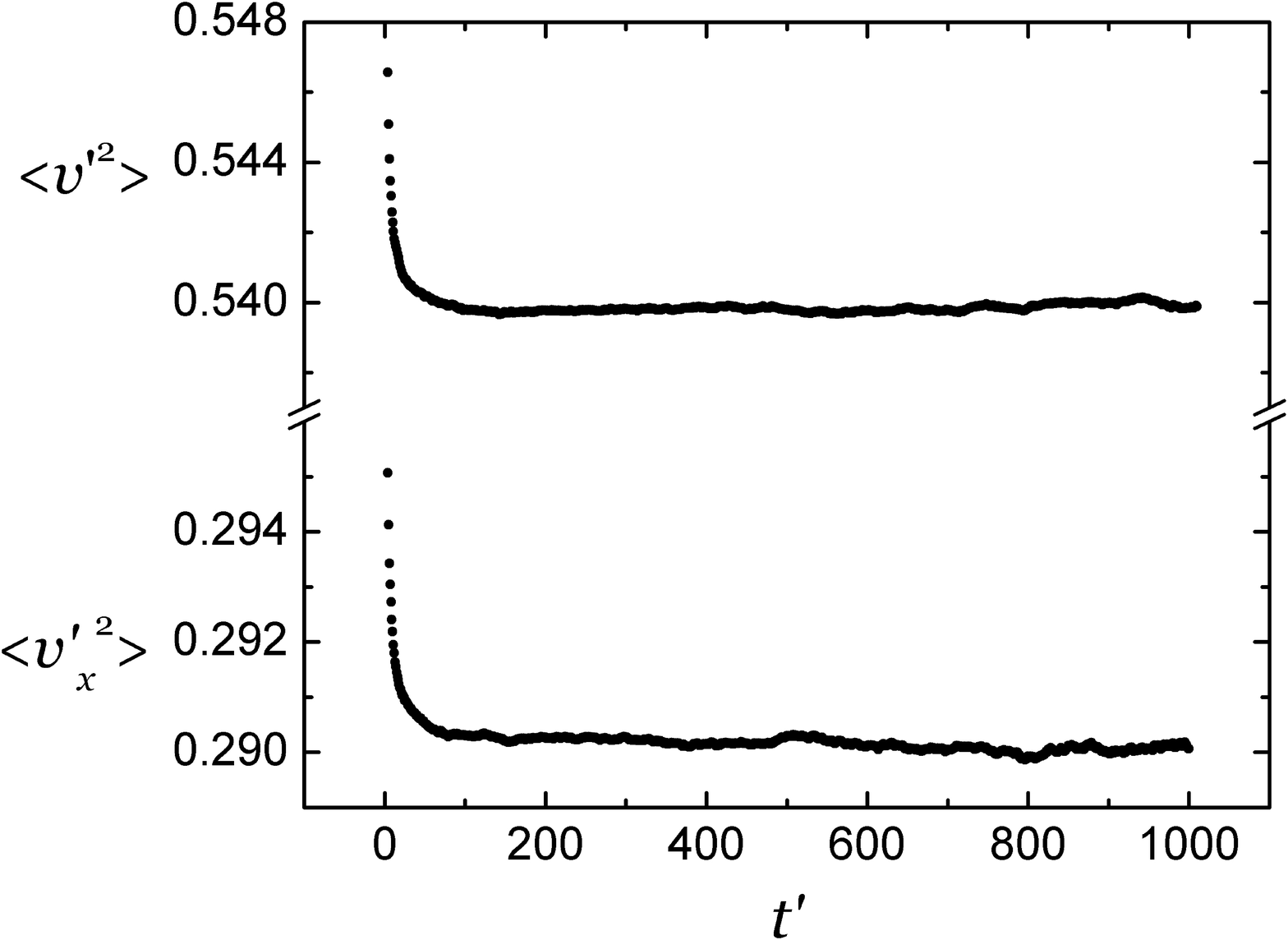}
\caption{$\langle v_x'^2\rangle^\vert_{{1\over 2},1}$ vs. $t'$ for a polymer on a
two-dimensional strip (lower curve) and $\langle\vec{v}\;'^2\rangle_{{1\over 2,}1}^\circ$ vs. $t'$ for a polymer in a three-dimensional channel with a circular cross section (upper curve).} \label{fig3}
\end{center}
\end{figure}

\newpage
\begin{figure}[Figure4]
\begin{center}
\includegraphics[width=18cm]{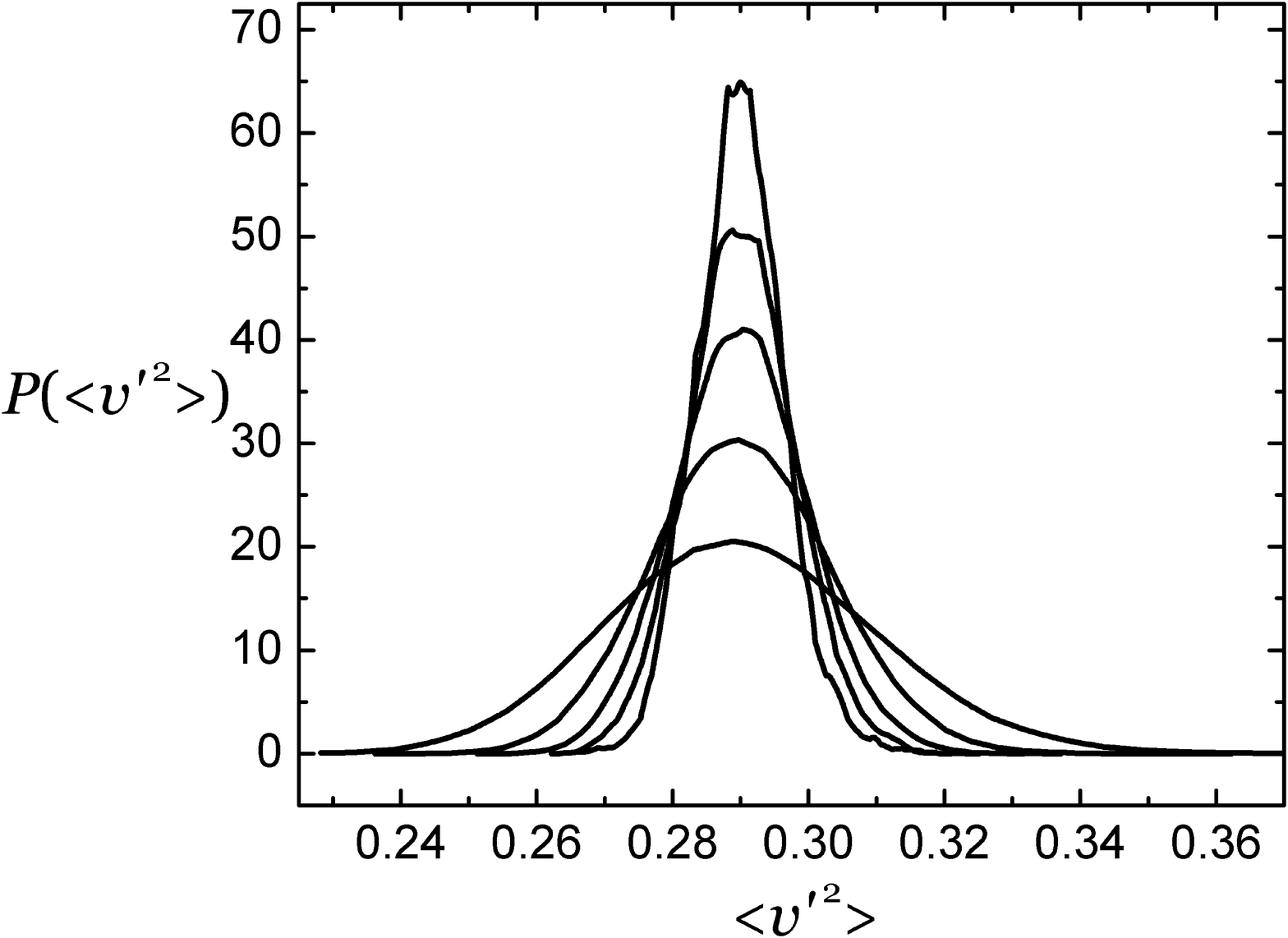}
\caption{Distribution $P(\langle v'^2\rangle)$ for a polymer on a two-dimensional strip, as defined below Eq. (\ref{alphaBoxcirc}). The curves correspond, from bottom to top, to $t'=$ 100, 225, 400, 625, and 900.}\label{fig4}
\end{center}
\end{figure}

\newpage
\begin{figure}[Figure5]
\begin{center}
\includegraphics[width=18cm]{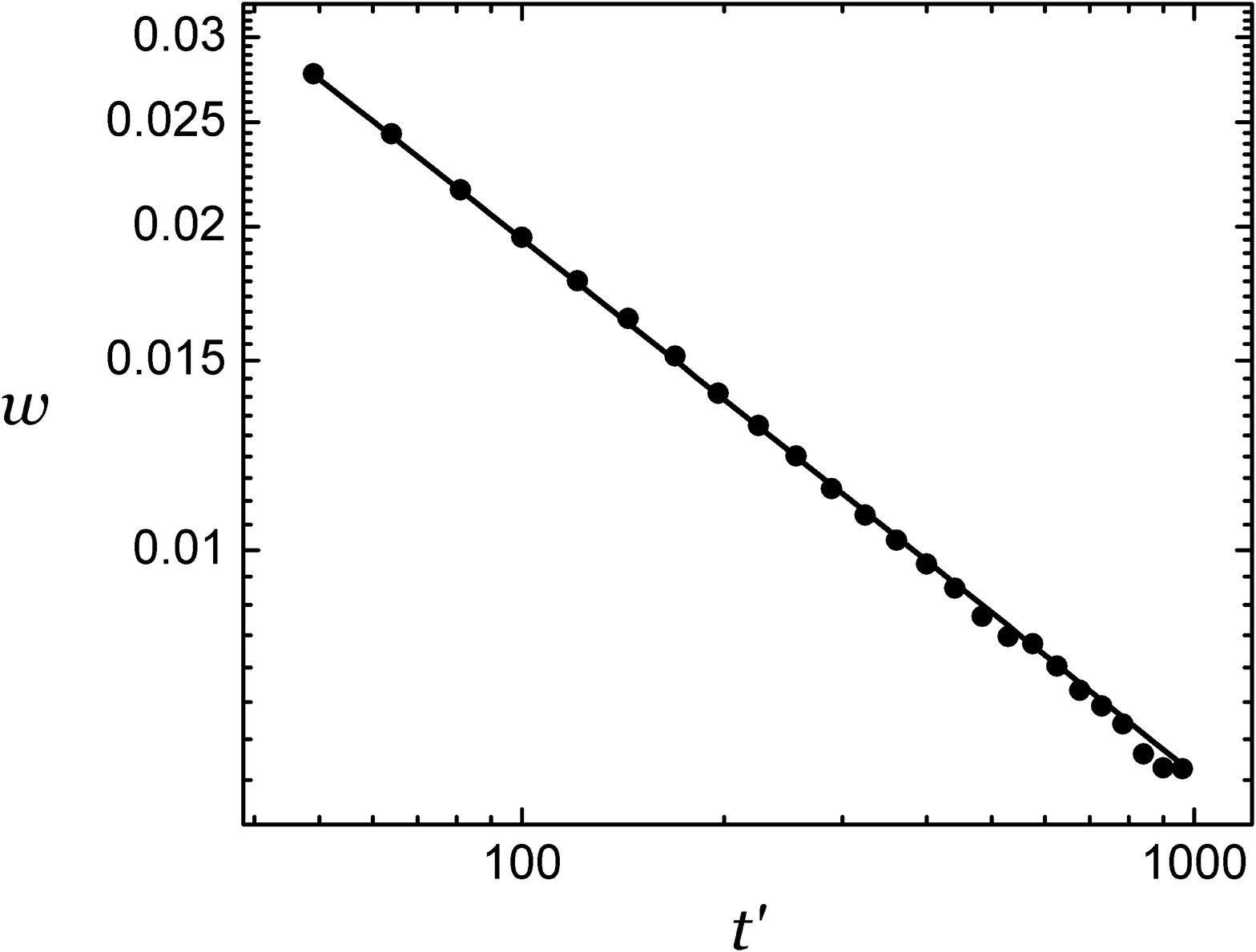}
\caption{Double logarithmic plot of the half width or standard deviation $w$ of the distribution $P(\langle v'^2\rangle)$ (see Fig. 4) as a function of $t'$. The straight line corresponds to $w=kt'^{-1/2}$, where $k$ is a constant.}\label{fig5}
\end{center}
\end{figure}
\end{document}